\documentclass[a4paper]{aa}
\usepackage{graphicx}
\begin{document}   
   \thesaurus{09.19.2, 13.09.3, 13.09.4}
\title{Element mixing in the Cassiopeia~A supernova\thanks{Based on
observations with ISO, an ESA project with instruments funded by ESA Member
States (especially the PI countries: France, Germany, the Netherlands and
the United Kingdom) and with the participation of ISAS and NASA, and on 
observations obtained at the Canada-France-Hawaii Telescope.}}
\author{T. Douvion\inst{1}, P.O. Lagage\inst{1} and C.J. Cesarsky\inst{1,2}}
   \offprints{tdouvion@cea.fr or lagage@cea.fr}
   \institute{\inst{1}CEA/DSM/DAPNIA/Service d'Astrophysique, CE Saclay, F-91191 Gif-sur-Yvette, France\\
\inst{2} European Southern Observatory, Karl Schwarzchild Str. 2, 85748 Garching bei Muenchen, Germany}

   \date{Received 7 September 1999 / Accepted 13 October 1999}

\maketitle
\markboth{T. Douvion et al.: Element mixing in the Cas~A supernova}{}
\begin{abstract}
Thanks to mid-infrared observations, we provide new clues to the element mixing during a 
supernova explosion by probing the mixing between three adjacent layers: the oxygen 
burning products layer (sulfur, argon,...), the silicate layer and the neon layer. The 
silicate and neon layers are both contaminated by sulfur and argon in a macroscopic 
way, but appear segregated, so that the mixing is heterogeneous. This finding complements 
the microscopic mixing information deduced from presolar grains found in meteorites and 
implies that, at present time, supernovae are probably not the main dust factory in 
the Galaxy. The mixing 
is often interpreted in terms of hydrodynamical instabilities driven by the outward shock 
following the implosion of the supernova core. Testing whether such instabilities can 
lead to the injection of material from a layer into upper layers without complete mixing, 
as suggested by the observations presented in this paper, should be possible
with the intense 
lasers which are starting to be used to simulate astrophysical plasmas.

\end{abstract}

\section{Introduction}
Supernovae (SNe) are key objects in the Universe (review by Trimble 1983 and
references therein). 
They are the factories which feed the interstellar medium 
with many of the heavy elements. These heavy elements are built up, layer 
by layer, inside massive stars, stratified according to the atomic number 
(review by Arnett 1995 and references therein). Freshly ejected supernova 
material can be directly observed
in the Cassiopeia~A supernova remnant. The Cassiopeia~A (Cas~A) SuperNova 
Remnant (SNR) is the youngest SNR known in our galaxy. The SuperNova (SN)
exploded about 320 years ago (Fesen, Becker and Goodrich 1988) at a distance of about 3.4 kpc 
(Reed et~al. 1995). It must have been subluminous and/or heavely obscured, 
since it passed unnoticed, except perhaps by Flamsteed in 1680 
(Ashworth 1980). The progenitor of the SN was a massive star 
(Vink, Kaastra and Bleeker 1996, Jansen et~al. 1988, Fabian et~al. 1980), probably of Wolf-Rayet type 
(Fesen, Becker and Goodrich 1988).\\
The freshly ejected SN material has been widely observed in the 
optical range (Baade and Minkowski 1954, Chevalier and Kirshner 1979,
van den Bergh and Kamper 1985 and earlier papers).
 The SN material is spatially distributed in Fast Moving Knots 
(FMKs) (see Figure 1) with a typical speed in the 5000\,km/s range, as deduced 
from their proper motion on the sky and from the Doppler shift of the lines 
emitted by them. The optical observations have revealed the presence in 
these knots of heavy elements such as oxygen, sulfur, argon, but no hydrogen 
or helium lines. Mid-InfraRed (Mid-IR) observations of the FMKs have only 
started recently thanks to observations with ISO, the Infrared Space 
Observatory (Kessler et~al. 1996); these observations have revealed the presence 
of two additional key components: neon and silicate dust (Lagage et~al. 1996, 
Arendt, Dwek and Moseley 1999).

In this paper, we present a new set of spectro-imaging observations made with 
ISOCAM (Cesarsky et~al. 1996), the camera on board of ISO; these observations reveal 
the spatial distribution of the silicate knots and of the neon knots. The comparison 
of these distributions brings unique information on the degree of mixing of the 
various elements which has occurred during the supernova explosion. Section 2 present 
the data, the data reduction and the results we obtain. In Section 3, we 
discuss the implication for the mixing in SNe.

\section{Observations and results}
   \begin{figure*}
\resizebox{1.0\hsize}{!}{\includegraphics[width=\hsize]{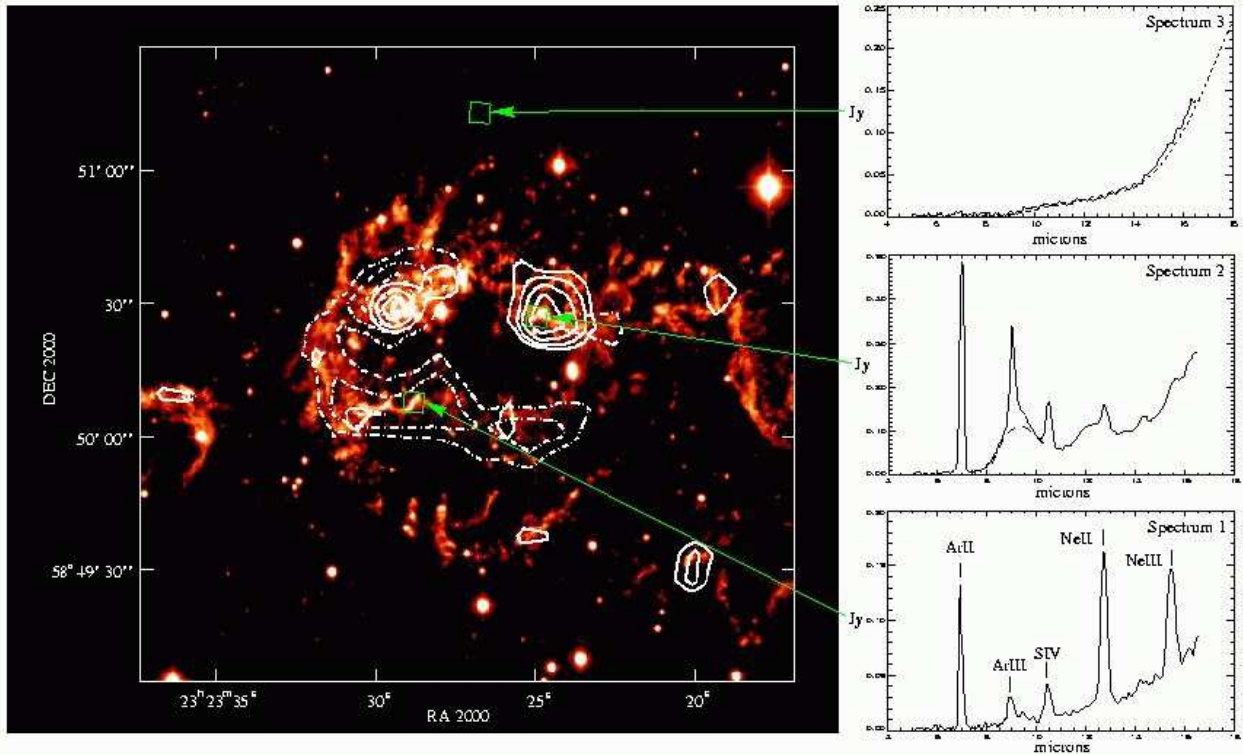}}
\caption[]{Contour maps of the [Ne~II] line emission (dotted contours) 
and of the 9.5\,$\mu$m silicate dust emission
 (full contours) overplotted onto an optical image of the Cassiopeia-A supernova remnant.
 At each contour level the flux is divided by a factor 1.5; (instrumental ghost effects were taken into 
account to limit the 
lower contours of neon). The neon line and silicate emission are deduced from spectra
such as those on the right of the image (see text). The spectra have been 
obtained with ISOCAM at a spatial resolution of 6''x6''; the ISOCAM pixel 
corresponding to the spectrum is represented on the
image by green squares. The optical image has been obtained with the SIS 
instrument at CFHT.}
         \label{Fig1}
    \end{figure*}

The observations were 
performed on December 5th 1996 with ISOCAM. 
The pixel field of view of the instrument 
was set to 6'', comparable to the diffraction limit of the telescope. 
The total field of view is 3'x3' and
the field was centered on the northern part of the remnant. For each pixel, 
we have a spectrum from 5 to 16.5 microns that was obtained by 
rotating the Circular Variable Filter of ISOCAM; the spectral resolution 
obtained this way is around 40. The data reduction was performed with
CIA\footnote{CIA is a joint development by the ESA 
astrophysics division and the ISOCAM consortium led by the ISOCAM PI, C.J. Cesarsky, 
Direction des sciences
de la mati\`ere, C.E.A., France.}, using a full spectroscopic data set of an 
off-position field to subtract the zodiacal contribution. The result
consists in 1024 spectra. Some of them are shown on Figure 1. They feature
both continuum emission and line emission. 
The continuum emission rises slowly from 8 to 16\,$\mu$m; in some spectra, a
bump is present around 9.3\,$\mu$m.\\
The lines are identified as [Ar~II] (7.0\,$\mu$m), [Ar~III] (9.0\,$\mu$m), 
[S~IV] (10.5\,$\mu$m), [Ne~II] (12.8\,$\mu$m) and [Ne~III] (15.5\,$\mu$m). The argon and 
sulfur lines where also observed with ISOPHOT (Tuffs et~al. 1997), but not the
neon lines, which are out of the ISOPHOT-S wavelength range.
The neon emission map is obtained by the difference between the peak 
flux in the [Ne~II] line and the underlying continuum. The evidence for 
the presence of dust is provided by the continuum radiation underlying 
the line emission; the silicate dust is well characterized by its 
feature around 9.3\,$\mu$m, (see spectrum 2 of Figure 1). 
The silicate emission map is obtained at 9.5\,$\mu$m by subtracting 
from the detected emission, the emission from a blackbody fitting 
the data at 7.5 and 11.5\,$\mu$m. These two maps are both overplotted on
an optical image obtained with the SIS instrument mounted on 
the Canada France Hawaii telescope on August 1998; 
a filter centered at 6750\,\AA$\:$ and with a band-pass of 780\,\AA$\:$ 
was used; the pixel field of view was 0.15\,arcsec and the integration time was
300\,s. Note that several optical knots are often present in a single ISOCAM pixel.\\

Neon, which is barely detected in the visible (Fesen 1990), gives prominent lines in 
the Mid-IR: the [Ne~II] line at 12.8\,$\mu$m and the [Ne~III] line at 15.5\,$\mu$m
(see spectrum~1 of Figure~1). Mid-IR searches of neon are more advantageous than
optical studies, because of their insensitivity to the relatively high interstellar extinction toward Cas~A 
(typically A$_{V}$=5, see Hurford and Fesen 1996) and of the lower temperature needed to excite IR 
lines compared to optical lines.
The neon map is compared with the silicate dust map on Figure 1.
Spectrum 1 is typical of neon knots. Spectrum 2 is typical of silicate
knots (the silicate feature at 9.3\,$\mu$m is underlined by the dotted 
curve). Spectrum 2 is slightly contamined by neon, certainly due to the 
strong neon emission just nearby. The bump around this small 
neon feature could be attributed to Al$_{2}$O$_{3}$ (Koike et~al. 1995, Kozasa and Hisoto 1997).\\
An anticorrelation between the presence of neon and the presence of silicate in many 
knots is evident. The regions where both neon and silicate are observed in the IR spectra 
are confused regions where several bright optical knots lie along the line of sight probed 
by an ISOCAM pixel (6''x6'').\\

The Mid-IR radiation can also be used to probe the presence of argon, through the [Ar~II] 
and [Ar~III] lines at respectively 7.0 and 9.0 microns, and of sulfur, through the [S~IV] 
line at 10.5 microns. The lines associated with these two elements show up in almost all 
the IR spectra, but, given that several knots lie in an ISOCAM pixel, additional arguments
are needed before claiming that knots emitting Neon in the IR also contain sulfur and argon. The first
argument comes from the IR data themselves. Indeed, even with the poor spectral 
resolution of ISOCAM observations, it has been 
possible to measure the Doppler shift of the lines emitted from the knots with the highest 
radial velocities (Lagage et~al. 1999). For these knots, all the lines have the same Doppler 
shift, indicating a common knot origin of neon, argon and sulfur.
Another way to find if neon, sulfur and argon originate
from the same knots is to search for oxygen, sulfur and argon lines in optical spectra.
Indeed in supernovae the neon layer is associated to the oxygen layer (see Figure~2) and
the optical [O~III] line has excitation conditions intermediate between those of the IR 
[Ne~II] and [Ne~III] lines (same critical density as the [Ne~II] line and ionization
potential intermediate between those of the [Ne~II] and [Ne~III] lines).
For this purpose we performed follow-up spectro-imaging observations 
of the Mid-IR knots in the optical at the Canada-France-Hawaii Telescope. Most of the FMK 
optical spectra feature oxygen, sulfur and argon lines, in agreement with previous studies 
(Hurford and Fesen 1996, Chevalier and Kirshner 1979). Thus we can conclude that argon and sulfur are 
indeed present in most of the neon knots. 
Note also that given that the [Ar~III] line and the [Ne~II] line have very similar
excitation conditions, we can exclude the presence of neon in the silicate knots 
which emit in the [Ar~III] line but not in the [Ne~II] line.\\

 Finally, spectrum 3 of Figure 1 originates in a region which is not associated with 
fast moving knots, but which nevertheless is bright in the IR (and also in X-ray and
radio).
No line emission is present in the spectrum, and the continuum emission is
well fitted by Draine and Lee silicates (Draine and Lee 1984) at a temperature of 105\,K
(see dashed line in spectrum 3 of figure 1); Draine and Lee graphites do not
fit this spectrum. The emission is probably due to circumstellar or interstellar
dust heated by the supernova blast wave.
Such a continuum emission is present all over the supernova remnant and 
is probably at the origin of the continuum dust emission
underlying the line emission in neon knots.
In spectrum 3, no room is left to synchrotron radiation down to the sensitivity limit 
of our observations, $2.10^{-3}$ Jansky (1\,$\sigma$) at 6 microns in an ISOCAM pixel. 
This limit is compatible with the expected IR synchrotron emission 
(about half a mJy), as extrapolated from the radio synchrotron emission between
 1.4\,Ghz and 4.8\,Ghz detected in the region of the ISOCAM pixel of
spectrum 3 (Anderson et~al. 1991).\\

\section{Discussion}
   \begin{figure*}
\resizebox{0.7\hsize}{!}{\includegraphics[width=0.7\hsize]{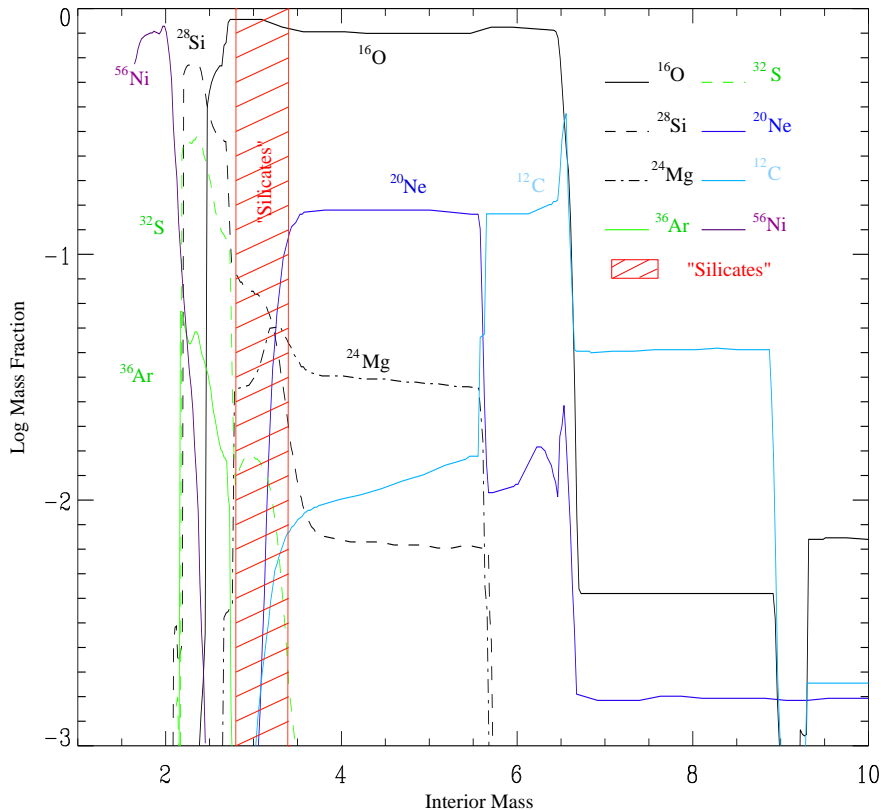}}
\hfill      \parbox[b]{5cm}{\vspace{-4cm}\caption[]{Layers of elements inside a 
supernova without mixing; (adapted from 
Woosley et Weaver (1995) for a 25 M$_{\odot}$ supernova progenitor; results are similar
for a 15 M$_{\odot}$ supernova progenitor). The hatched region 
is the "silicate" region (see the text).\\
\\
}}
         \label{Fig2}
;\vspace{-1cm}
    \end{figure*}

In order to explain these observations, we have to recall how the elements 
are structured inside a supernova. The elements are located in a stratified 
way according to their burning stage. Neon, silicate making elements and 
oxygen burning products (S, Ar...) are in different layers (see Figure 2). 
The hatched region on Figure 2 is what we call the "silicate" region; 
this is where sufficient amounts of
oxygen, magnesium and silicon are present at the same time, 
to form pyroxene (MgSiO$_{3}$).
We consider pyroxene because it is this type of silicate which is 
predicted by dust formation models
(Kozasa, Hasegawa and Nomoto 1991). Furthermore the observation of a 22\,$\mu$m feature in 
Cassiopeia~A was attributed to Mg protosilicates (Arendt, Dwek and Mosely 1999)
and the 9.3\,$\mu$m feature of our spectra (see spectrum 2 of Figure 1)
is well fitted with very small grains (less than 0.1 microns) of pyroxene.\\
Then a straightforward way to interpret the fact that the neon and silicate layers remain 
spatially anticorrelated in the ejecta is to consider that there has only been weak mixing 
between those layers during the supernova explosion. In contrast, sulfur and argon 
have been extensively mixed. The mixing of the sulfur and argon layer with the oxygen 
layer was already revealed by optical observations; the mixing is known to be extensive, 
but not complete as some oxygen knots are free from sulfur and some sulfur knots
are free from oxygen (Chevalier and Kirshner 1979, van den Bergh and Kamper 1985). The key new result from the Mid-IR 
observations is that the neon and silicate layers are essentially unmixed, i.e. the mixing 
is heterogeneous.

The data also suggest that the mixing is mostly macroscopic. Indeed silicon is located 
in the same layer as sulfur and argon; thus silicon is also expected to be 
spread into the neon layer. If the silicon mixing were microscopic, silicates could be 
produced all-over the magnesium layer, which encompasses the neon layer (see Figure 2); 
then the anticorrelation of Figure~1 would not hold. Another possibility would be that the 
mixing is indeed microscopic, but that for some reasons, the silicate production is 
quenched in the neon layer. In any case the conclusion is that the silicate production 
from SN is limited to the thin layer shown in Figure 2. As a consequence, the silicate 
dust production is five times lower than in the case of condensation of all the silicate 
condensable elements. Silicates are known to be present in large quantities in the 
interstellar medium, but there is still a debate as to the dominant injection source of 
these silicates. Supernova ejecta or stellar wind material from giant and
supergiant stars have been invoked. Core collapse SNe could be the main silicate provider in case of complete 
condensation of silicate elements (Dwek 1998). But, if for all core collapse SNe, the 
condensation is 
incomplete, as discussed here in the case of Cas~A, then core collapse SNe can no
longer be the dominant source of silicates at present; but they could
have played a dominant role in the past (Dwek 1998). No evidence for dust formation
in type Ia SNe has been found so far and no information on the mixing of silicate elements
in those SNe exist. 
In the absence of this evidence, we consider that giant and supergiant 
stars are likely to be the main silicate providers.

The origin of the mixing is still an open question. Several studies have been made in 
the framework of the observations of SN 1987A (Arnett et~al. 1989 and references
therein), 
the supernova whose explosion in the Large Magellanic Cloud was detected 13 years ago, and of SN 1993J 
(Spyromilio 1994, Wang and Hu 1994); the issue is how to mix the inner regions where the nickel 
and cobalt have been synthetized by oxygen explosive burning (see Figure 2) with the 
upper layers of hydrogen and helium. Hydrodynamic instabilities (of the Rayleigh-Taylor 
and Richtmyer-Meshkov type) are usually considered as playing a key role. Presupernova 
models show that the density profile of the presupernova features steep density gradients 
at the interface of composition changes, especially at the interfaces hydrogen/helium and 
helium/oxygen; these gradients are regions where shock induced Richtmyer-Meshkov 
instabilities, followed by Rayleigh-Taylor instabilities, can develop during the SN explosion. Models 
based on instabilities at these interfaces meet with difficulties to reproduce 
quantitatively the data (Arnett 1995). It is also hard to imagine how instabilities at 
the H/He or the He/O interface could be responsible for the heterogeneous mixing 
presented here, especially if, as generally claimed, the Cas~A progenitor has already 
shed all of its hydrogen and all or most of its helium before the SN exploded. 
Presupernova models feature another, weaker, density gradient at the bottom of the oxygen 
layer (for example Nomoto et~al. 1997). In addition, convection at work in this region 
during the presupernova phase can generate density perturbations which could seed the 
instabilities (Bazan and Arnett 1998). Thus it seems likely that the mixing originated at the bottom 
of the oxygen layer, but this remains to be proven by self consistent numerical models 
following up all phases from pre-supernova to now, taking into account radiative cooling 
which can lead to clumps.

In complement to advances in numerical simulations, laboratory experiments are needed. 
Such experiments are starting to be possible thanks to the use of intense lasers, which 
can generate plasmas mimicking various astrophysical conditions (Remington et~al. 1999). 
Laboratory experiments simulating hydrodynamic instabilities at the H/He interface have 
already been conducted (Kane et~al. 1997, Drake et~al. 1998). Experiments reproducing the 
conditions at the oxygen/oxygen-burning product interface should be performed. The 
possibility of heterogeneous mixing could be tested.

Another issue which should be investigated is the degree of microscopic mixing versus
macroscopic mixing. Heterogeneous microscopic mixing in supernovae is a key requirement in order 
to explain the isotopic anomalies observed in some presolar grains found in meteorites 
(Travaglio et~al. 1998). For example, the presence of silicon carbide with $^{28}$Si implies 
that $^{28}$Si, produced in an inner shell of a SN, has to be injected up to the outer 
layer of carbon and then microscopically mixed with the carbon. The mixing has to be 
heterogeneous in the sense that the oxygen layer should not be mixed with the carbon 
layer; otherwise, the carbon would be locked into CO molecules and no carbon dust 
particle could be made. Rayleigh-Taylor instabilities mostly lead to macroscopic mixing, 
but some microscopic mixing could occur at the interface of macroscopically mixed 
regions. In that context, Mid-IR observations presented here are complementary to 
meteorite studies. \\

\bigskip {\bf Acknowledgment} We would like to thank J.P. Chieze and R. Teyssier for 
enlightening discussions about Rayleigh-Taylor instabilities and laser experiments. We 
thank the referee R. Arendt for his careful reading of the manuscript and his 
useful comments.

{}
\end{document}